\documentclass[manuscript]{aastex}

\slugcomment{Submitted to ApJ}

\shorttitle{Perpendicular Transport of Solar Energetic Particles}
\shortauthors{Strauss and Fichtner}

\begin{document}


\title{On Aspects Pertaining to the Perpendicular Diffusion of Solar Energetic Particles}


\author{R. D. Strauss\altaffilmark{1,2} and H. Fichtner\altaffilmark{2}}


\altaffiltext{1}{Centre for Space Research, North-West University, South Africa.}
\altaffiltext{2}{Institut f\"ur Theoretische Physik IV, Ruhr-Universit\"at Bochum, Germany.}

\begin{abstract}

The multitude of recent multi-point spacecraft observations of solar energetic particle (SEP) events have made it possible to study the longitudinal distribution of SEPs in great detail. SEPs, even those accelerated during impulsive events, show a much wider than expected longitudinal extent, bringing into question the processes responsible for their transport perpendicular to the local magnetic field. In this paper we examine some aspects of perpendicular transport by including perpendicular diffusion into a numerical SEP transport model that simulates the propagation of impulsively accelerated SEP electrons in the ecliptic plane. We find that: (i) The pitch-angle dependence of the perpendicular diffusion coefficient is an important, and currently mainly overlooked, transport parameter. (ii) SEP intensities are generally asymmetric in longitude, being enhanced towards the west of optimal magnetic connection to the acceleration region. (iii) The maximum SEP intensity may also be shifted (parameter dependently) away from the longitude of best magnetic connectivity at 1 AU. We also calculate the maximum intensity, the time of maximum intensity, the onset time and the maximum anisotropy as a function of longitude at Earth's orbit and compare the results, in a qualitative fashion, to recent spacecraft observations.

\end{abstract}


\section{Introduction}

With the launch of the twin STEREO spacecraft, it is possible to observe solar energetic particle (SEP) events simultaneously by means of {\it in-situ} particle observations \citep[e.g.][]{dresingetal2012} and remote sensing observations of the associated acceleration regions \citep[e.g.][]{klassenetal2012}. Recent observations by, e.g., \citet{dresingetal2012} and \citet{drogeteal2014} have shown that, even for impulsive SEP events, the longitudinal spreads of the particles of a given event is much wider that previously thought, even extending to almost $360^{\circ}$ in longitude at 1 AU. It is still undecided what process is primarily responsible for the longitudinal transport of SEPs during these wide spread observations, with three main theories (explanations) put forward: (i) Effective diffusion perpendicular to the mean Parker \citep[][]{parkerhmf} heliospheric magnetic field \citep[HMF, e.g.][]{drogeetal2010}, (ii) a changing HMF topology, possible due to the passage of a coronal mass ejection, that can enhance the longitudinal transport of SEPs \citep[e.g.][]{tanetal29012,tan2009} and (iii) an extended source close to the Sun formed by, e.g., effective azimuthal particle transport in the corona \citep[][]{dressing2014}. In reality, it is likely a combination of these processes that contribute to the longitudinal transport of SEPs, although it is uncertain which is dominant. In this paper, we present model solutions of simulated impulsive SEP events in the ecliptic plane of the heliosphere, where perpendicular diffusion of SEPs in a Parker HMF geometry is included. We focus on simulating observable SEP quantities as a function of longitude along the Earth's orbit. These simulation can, in future, be compared directly to observations and may then be used to determine to what extent perpendicular diffusion is the dominant transport process of SEPs.
\section{The Transport Model}

In this work, we consider the transport of SEP electrons, accelerated impulsively near the Sun \citep[see, e.g.,][for a review]{remaesreview}. These particles have assumed energies of $E\sim 85$ keV and a corresponding speed of $v\sim 3.7$ AU.hr$^{-1}$, so that, solar wind effects (both convection and adiabatic energy losses) may be neglected \citep[e.g.][]{ruffolo1995}. The relevant transport equation (TPE) for these particles is therefore \citep[e.g.][]{skilling1971,sclikkie2002}

\begin{eqnarray}
\label{Eq:TPE}
\frac{\partial f(\vec{x},\mu,t)}{\partial t} = - \nabla \cdot \left( \mu v \hat{b} f \right) - \frac{\partial}{\partial \mu} \left( \frac{1-\mu^2}{2L} vf \right) \nonumber \\ 
+ \frac{\partial}{\partial \mu} \left(D_{\mu\mu} (\vec{x},\mu) \frac{\partial f}{\partial \mu} \right)  +   \nabla \cdot \left( \mathbf{D}^{(x)}_{\perp}(\vec{x},\mu) \cdot \nabla f \right) \nonumber
\end{eqnarray}

and is, in this work, solved numerically in the ecliptic plane of the heliosphere {\bf(in terms of radial distance, or heliocentric distance, $r$ and azimuthal angle, or heliographic longitude, $\phi$)}; see also the work by \citet{zhangetal2009} and \citet{drogeetal2010}. This equation describes the evolution of a gyrotropic distribution of SEPs under the influence of the following processes (described by the terms on the right): Particle streaming along the mean heliospheric magnetic field (HMF), focusing in the diverging HMF, pitch-angle scattering and diffusion perpendicular to the mean HMF. In these expressions, $v_{||} = \mu v$ is the parallel (to the HMF) speed and $\mu$ the cosine of the particle pitch-angle. The unit vector along the HMF is indicated by $\hat{b}$, while the focusing length is defined as

\begin{equation}
L^{-1} = \nabla \cdot \hat{b}.
\end{equation}

{\bf A calculation of $L$, for a Parker HMF, is presented by \citet{hewang2012}}. For the pitch-angle diffusion coefficient $D_{\mu\mu}$, we adopt the form

\begin{equation}
D_{\mu\mu}(r,\mu,\phi) = D_{\mu\mu,0}(r,\phi) \left( 1 - \mu^2 \right) \left\{ \left| \mu \right|^{q-1} + H \right\}
\end{equation}

used by e.g. \citet{drogeetal2010}. Here, $q=5/3$ is the spectral index of the inertial (Kolmogorov) range of the turbulent power spectrum and $H = 0.05$, although chosen in an {\it ad hoc} fashion, allows for the presence of non-linear effects \citep[e.g.][]{shalchi2005}. Following the standard definition \citep[][]{hasselmaanwibberenz1968} of the parallel mean free path,

\begin{equation}
\lambda_{||}(r,\phi) = \frac{3v}{8} \int_{-1}^{+1} \frac{\left( 1 - \mu^2 \right)^2}{D_{\mu\mu}(r,\phi,\mu)} d\mu \nonumber
\end{equation}

the value of $D_{\mu\mu,0}$ is not specified directly in the transport model, but rather the value of $\lambda_{||}$, from which $D_{\mu\mu,0}$ may be calculated via the equation above. Note that, in Eq. \ref{Eq:TPE}, $D_{\perp}^{(x)}$ must be specified in the global coordinate system used to solve the TPE. Such a transformation from HMF aligned coordinates to spherical spatial coordinates, adopted in this study, is described in Appendix \ref{Sec:appen1}. For simplicity, it is assumed that the transport parameters have no azimuthal dependence.

Once that Eq. \ref{Eq:TPE} is solved to obtain $f$, we may also calculate the omni-directional intensity

\begin{equation}
F(r,\phi,t) =\frac{1}{2} \int_{-1}^{+1}   f(r,\phi,\mu,t)  d\mu
\end{equation}

and the first order anisotropy

\begin{equation}
A(r,\phi,t) = 3 \frac{\int_{-1}^{+1}  \mu f  d\mu}{\int_{-1}^{+1} f   d\mu} \nonumber
\end{equation}

which can be compared directly to observations.

As a boundary condition, the following isotropic injection function

\begin{equation}
f(r,\phi,t) =g(t)
  \cdot \exp \left[ - \frac{(\phi - \phi_0)^2}{\phi^2_m}  \right] \cdot \delta (r - r_0) 
  \end{equation}

is prescribed at the inner boundary, located at $r_0 = 0.05$ AU. Gaussian injection in $\phi$ is assumed with $\phi_0 = \pi/2$ and $\phi^2_m = 0.05$ rad$^2$. The form of this injection function is, of course, not well known and continues to be refined by means of simulations \citep[e.g.][]{heetal2011}. Because the source is assumed to have a finite azimuthal size, lateral particle transport close to the Sun is implicitly assumed in the model. By changing the azimuthal extent of the source region (a topic not discussed in this paper), we may, in an {\it ad hoc} fashion also change the effectiveness of azimuthal transport at $r<r_0$. The temporal dependence of the injection is described by $g(t)$, which is discussed later.

\section{The Functional Form of the Perpendicular Diffusion Coefficient}

In this section, we challenge the statement of e.g. \citet{qinetal2013} that the pitch-angle dependence of $D_{\perp}$ is not an important parameter to consider when modelling SEP transport. This statement is of course true when $f$ is nearly isotropic (i.e., when pitch-angle diffusion is extremely efficient), but under normal propagation conditions, effective focusing near the Sun produces large anisotropies at Earth, and consequently, the functional form of $D_{\perp}$ can be very important. 

We consider three different forms of $D_{\perp}$ that are widely assumed in the SEP transport literature, namely: (i) When $D_{\perp}$ is independent of $\mu$

\begin{equation}
D^{Constant}_{\perp}(\mu) = D_{\perp,0}
\end{equation}

(ii) The well known field line random walk \citep[FLRW,][]{jokipii1966,qinshalchi} coefficient

\begin{equation}
D^{FLRW}_{\perp}(\mu) = 2 D_{\perp,0} \left| \mu  \right|
\end{equation}

and lastly, (iii) the phenomenological form proposed by \citet{drogeetal2010}

\begin{equation}
D^{Scattering}_{\perp}(\mu) = \frac{4}{\pi} D_{\perp,0}\sqrt{ 1 - \mu^2  }.
\end{equation}

\begin{figure}
\epsscale{.50}
\plotone{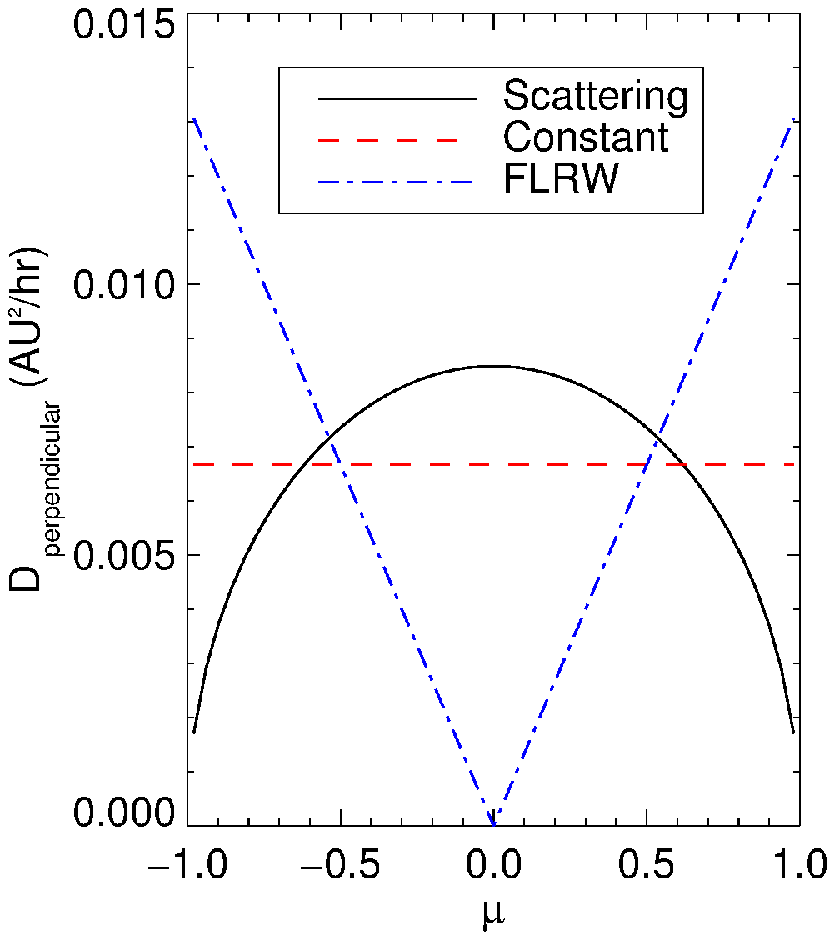}
\caption{{The different functional forms of $D_{\perp}$ considered in this study. Note that all of these choices lead to the same value of $\kappa_{\perp}$ when averaged over pitch-angle}.\label{fig1}}
\end{figure}

This last choice is motivated by the assumption that $D_{\perp}$ should generally increase with the particles' Larmor radius, and as such, should scale as $D_{\perp} \sim v_{\perp}$, where $v_{\perp}=v\sqrt{1 - \mu^2}$. The normalization factors in the equations above (i.e. the constants in front of $D_{\perp,0}$) above were chosen such that when the isotropic perpendicular diffusion coefficient

\begin{equation}
\kappa_{\perp}(r)  = \frac{1}{2}\int_{-1}^{+1} D_{\perp}(r,\mu) d\mu
\end{equation}

is calculated, all of these different forms lead to the same value of $\kappa_{\perp}(r)  = D_{\perp,0}(r)$. Calculating the corresponding perpendicular mean free path, $\lambda_{\perp}=3\kappa_{\perp}/v$, it is furthermore assumed that $\lambda_{\perp}=\eta \lambda_{||}$ with $\eta$ a constant. The different function forms of $D_{\perp}$ considered in this study is shown in Fig. \ref{fig1} as a function of $\mu$.

\begin{figure}
\epsscale{.750}
\plotone{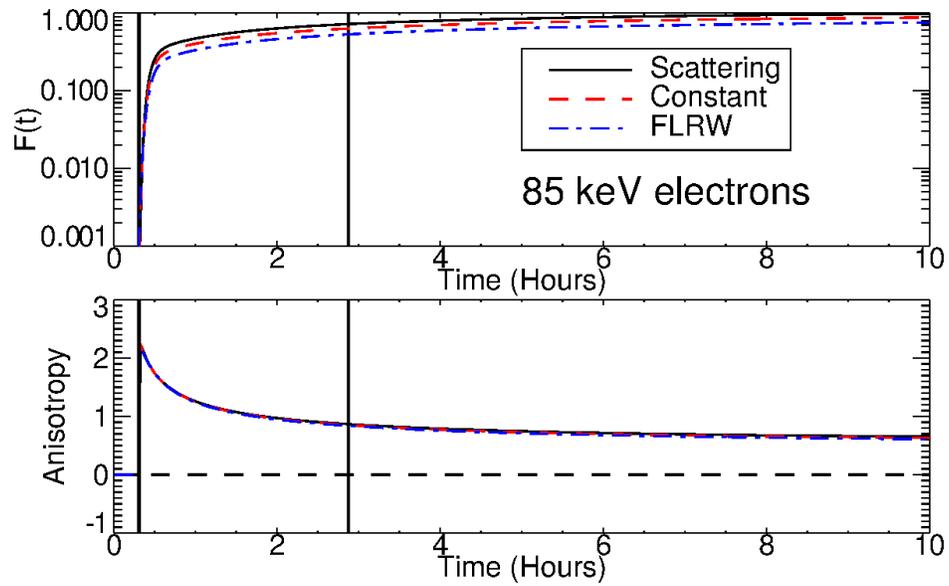}
\caption{{The temporal behaviour of the omni-directional intensity (top panel) and the anisotropy (bottom panel) at $r=1$ AU and at the azimuthal angle of best magnetic connectivity to the source}.\label{fig2}}
\end{figure}

For the results shown in this section, we assumed $\lambda_{||}=1$ AU (independent of spatial position) and $\eta=0.02$, along with a time-independent injection function, $g(t)=1$. Fig. \ref{fig2} shows the resulting omni-directional intensity (top panel) and anisotropy (bottom panel), as a function of time, for the three choices of $D_{\perp}$. The results are shown at Earth orbit (that is, at a radial distance of $r=1$ AU) and at the azimuthal angle of best magnetic connection to the source at the Sun. 

The two vertical lines on this graph (at $t \sim 0.3$ hrs and $t\sim 2.8$ hrs) show two limits of the model: (i) The first time scale is the minimum time an SEP, with this energy, needs to stream from the Sun to Earth orbit along the HMF (a distance of $\sim 1.2$ AU). Because no particle (should) generally reach the Earth before this time, it may be referred to as the {\it causality} time scale. (ii) The latter temporal limitation is due to our choice of placing the outer radial boundary at $r=3$ AU. This time scale is the time needed for a SEP to stream from the Sun, up to the model boundary, and back to Earth (a distance of $\sim 10.4$ AU). Although not a significant effect, the model solutions beyond this time may contain boundary condition effects. Looking at Fig. \ref{fig2}, it is clear that the different choices of $D_{\perp}$ do not affect the temporal profiles at this position very significantly, as both the intensity and anisotropy (at least near the point of best magnetic connectivity) are rather governed by the interplay between pitch-angle diffusion and focusing.

\begin{figure}
\epsscale{.50}
\plotone{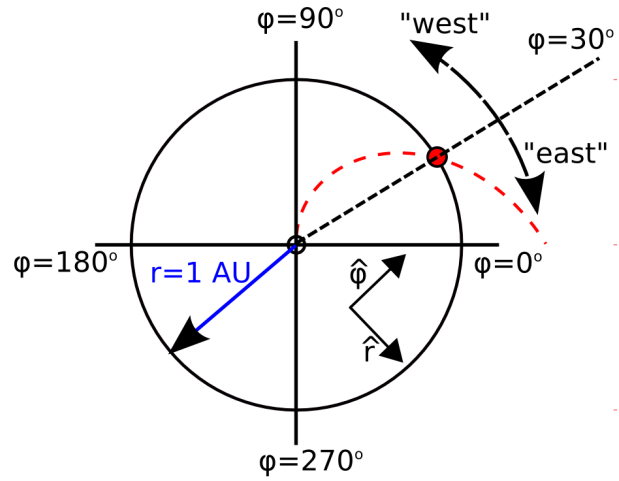}
\caption{{An illustration of the coordinate system used in this study. See the text for details}.\label{fig9}}
\end{figure}

Before examining the azimuthal dependence of the modelled particle intensities, it is useful to briefly define the coordinate system wherein the simulations are performed. Fig. \ref{fig9} shows a projection of a HMF line, connected to the assumed source at the Sun, onto the ecliptic plane (red dashed line). The azimuthal angle is defined to increase in the direction of solar rotation, i.e. counter-clockwise. In the simulations shown in this paper, the source (or more specifically, the maximum of the source) is assumed to be located at $\phi=90^{\circ}$ (the position from which the sketched HMF line originates). An observer, situated at Earth's orbit, would therefore be most optimally magnetically connected to the source at an angle of $\phi \sim 30^{\circ}$ (the red dot in the figure). With respect to this observer, increasing values of $\phi$ defines ``west of best magnetic connectivity'', while decreasing values define ``east''.

\begin{figure}
\epsscale{.50}
\plotone{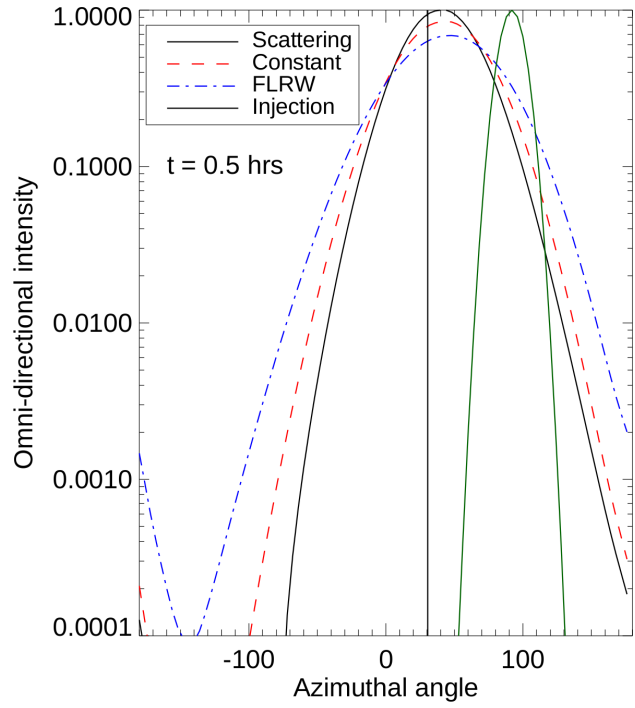}
\caption{{The calculated omni-directional intensity as a function of azimuthal angle at $t=0.5$ hrs for the different choices of $D_{\perp}$. The green Gaussian curve shows the injection function, while the vertical black line shows the azimuthal position of best magnetic connectivity to the source at 1 AU}.\label{fig3}}
\end{figure}

In Fig. \ref{fig3} we show the intensity (short for omni-directional intensity) as a function of $\phi$ at $t=0.5$ hrs. In this figure, the green curve shows the injected Gaussian distribution at the inner boundary, while the vertical line shows the angle of optimal magnetic connection at $1$ AU to the maximum value of the source. Here, the effect of the different diffusion coefficients are more evident, even for this relatively low value of $\eta=0.02$. The FLRW coefficient leads to the most efficient perpendicular diffusion, while the {\it scattering} coefficient (that is, $D_{\perp} \sim v_{\perp}$) is the most ineffective. This is because the SEP distribution between the Sun and Earth is generally highly anisotropic because of effective focusing, and the FLRW coefficient reaches it maximal value at $\mu=\pm 1$. In the FLRW limit therefore, a highly anisotropic beam of SEPs are most effectively scattered perpendicular to the mean field. 

\begin{figure}
\epsscale{.99}
\plotone{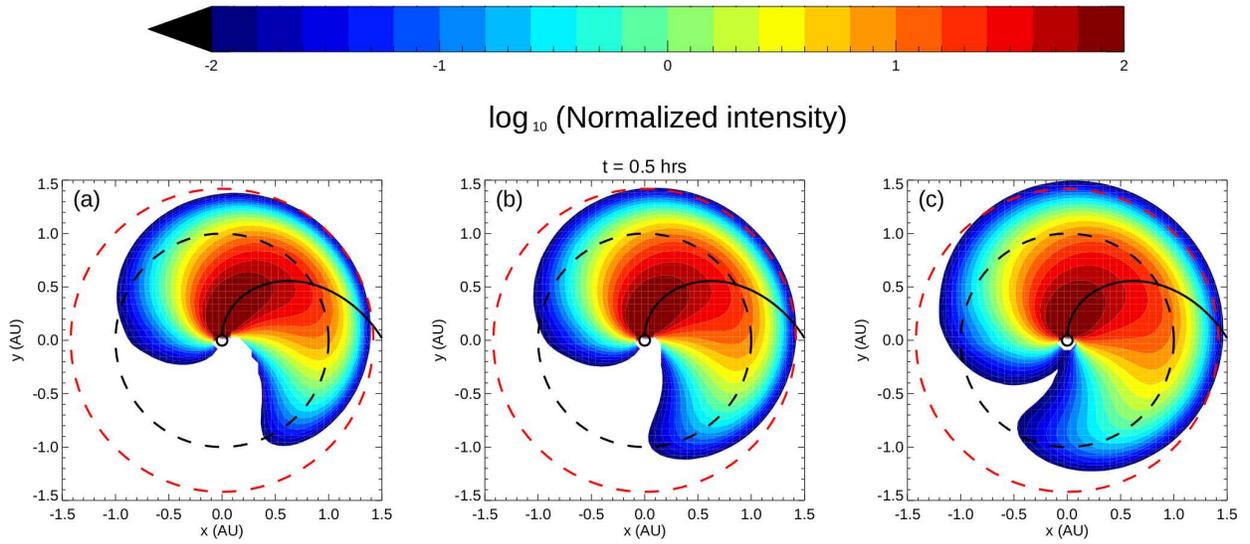}
\caption{{The omni-directional intensity in the ecliptic plane, at $t=0.5$ hrs, for the different choices of $D_{\perp}$: Panel (a) is for the case of $D_{\perp} \propto v_{\perp}$, panel (b) for the constant coefficient and panel (c) for the case of the FLRW coefficient. The solid black line shows the HMF line optimally connected to the source, while the dashed black and red lines show the trajectory of Earth and the causality requirement, respectively}.\label{fig4}}
\end{figure}

In Fig. \ref{fig4}, we show the intensity in the ecliptic plane, at $t=0.5$ hrs, for the three choices of $D_{\perp}$: Panel (a) for $D_{\perp}\sim v_{\perp}$, panel (b) for the case when $D_{\perp}$ is independent of $\mu$ and panel (c) the FLRW coefficient. The black dashed curve indicates Earth's orbit, while the red dashed curve shows the causality constraint (the maximum distance that an SEP may propagate along the HMF since injection). Similar to Fig. \ref{fig3}, we again note that the different choices of $D_{\perp}$ lead to different efficiencies of perpendicular diffusion and very different azimuthal SEP distributions. An interesting observations, discussed in Section \ref{Sec:discuss}, is that both panels (b) and (c) show particles beyond the causality limit.

\section{Symmetries Associated with SEP Transport}

\begin{figure}
\epsscale{.5}
\plotone{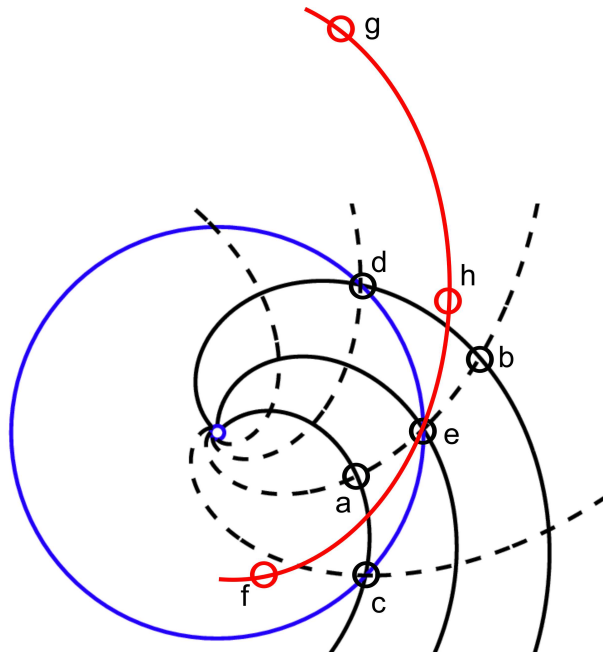}
\caption{{Testing the symmetry of a SEP distribution under the influence of both particle streaming and perpendicular diffusion. See the text for details}.\label{fig5}}
\end{figure}

The recent observations compiled by \citet{larioetal2013} and \citet{dressing2014} have brought into question the symmetrical nature, in terms of longitude or azimuthal angle, of the SEP distribution at 1 AU, and this topic is addressed in the following two sections. Fig. \ref{fig5} illustrates the problem of finding a suitable plane of symmetry for SEPs under the influence of particle streaming and perpendicular diffusion in the ecliptic plane. Assuming a point-like (or Gaussian in terms of $\phi$) injection of SEPs near the Sun, what type of distribution will a fleet of observers at 1 AU (the blue circle in the figure) measure? If only streaming is considered, SEPs would simply follow HMF lines (the solid black lines) and the observed SEP distribution would be symmetrical as measured along the line c-e-d (i.e. along a spherical orbit of constant HMF length). With symmetry we mean that the distribution will take the form of an azimuthally symmetrical Gaussian distribution, peaking at e (the point of optimal magnetic connection). Perpendicular diffusion however acts, as the term suggests, perpendicular to the field along the dashed lines shown in the figure. If this type of diffusion would dominate the transport process, the distribution would be symmetrical along the line a-e-b (i.e. perpendicular to the HMF). In reality, streaming and diffusion will compete with each other, so that the resulting distribution would rather be symmetrical in the line f-e-g. Moreover, because perpendicular diffusion also operates in the radial direction (away from the Sun if west of best connection and towards the Sun towards smaller values of $\phi$, i.e. towards the east) the distribution would not be a symmetrical Gaussian and will be enhanced towards the west (see also the results presented in Fig. \ref{fig3}) It is important to note that, as the HMF spiral angle reaches the limit of $\Psi \rightarrow 90^{\circ}$ (large radial distances), the HMF becomes essentially azimuthal, so that perpendicular diffusion (in this limit) leads to diffusion only in the radial direction. In the limit $\Psi \rightarrow 0^{\circ}$ (near the Sun), perpendicular diffusion acts purely in the azimuthal direction. An additional effect comes into play when $D_{\perp}^{(x)}$ is not constant: The so-called drift terms, $\nabla \cdot \mathbf{D}_{\perp}^{(x)}$, can also {\it convect} the distribution to either larger or smaller $r$ or $\phi$ values, depending on the sign of these derivatives. The peak of the SEP distribution may, therefore, be shifted away from the point of best magnetic connection, and may, for an illustrating example occur at point h in Fig. \ref{fig5}.

These effects are illustrated in this section, by using $\lambda_{||}=0.5$ AU and $\eta=0.02$ (although this value is changed later on). The injected SEP distribution follows a Reid-Axford temporal profile \citep[][]{reid1964} with

\begin{equation}
g(t)=\frac{C}{t} \exp \left[ -\frac{\tau_a}{t}  -\frac{t}{\tau_e}  \right]
\end{equation}

where $\tau_a = 1/10$ hr and $\tau_e = 1$ hr {\bf (the so-called acceleration and escape time scales for SEP acceleration and release from an active region)} and $C$ is a constant. Also note that the $D_{\perp}\sim v_{\perp}$ perpendicular diffusion coefficient is used for the rest of the study.

\begin{figure}
\epsscale{.75}
\plotone{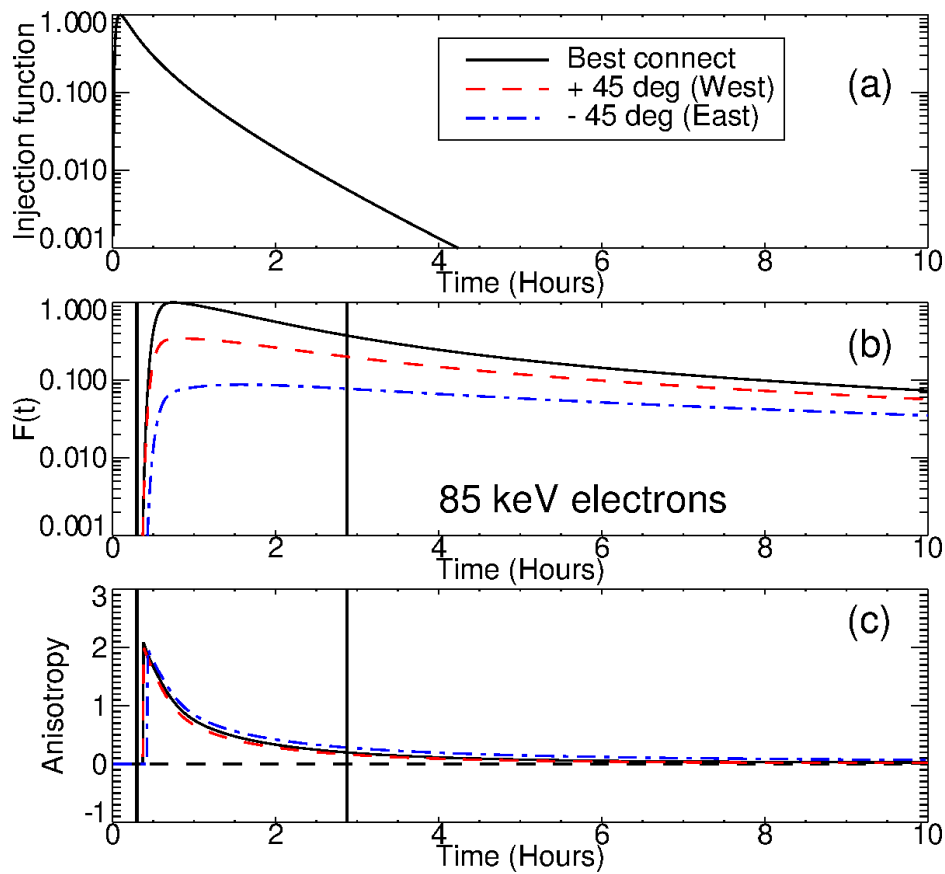}
\caption{{The top panel shows the assumed injection function, while panels (b) and (c) displays the resulting omni-directional intensity and anisotropy, as a function of time, at $r=1$ AU. The solutions are shown at an angle of optimal magnetic connection and two points $\pm 45^{\circ}$ away from it, as indicated in the legend}.\label{fig6}}
\end{figure}

Fig. \ref{fig6} shows the assumed injection function (panel a), the calculated intensity (panel b) and anisotropy (panel c) as a function of time at 1 AU. Three solutions are shown, corresponding to different azimuthal positions: Optimal magnetic connection (solid black lines) and two points $\pm 45^{\circ}$ away from it (dashed red and dash-dotted blue lines ,respectively). The middle panel illustrates the fact that the intensity is not symmetric about the point of best magnetic connection, with the flux enhanced towards the west (larger values of $\phi$), as compared to an equivalent point towards the east (a negative shift in $\phi$). The behaviour of the anisotropy is discussed in the next section, but is generally anti-correlated with the intensity.

\begin{figure}
\epsscale{.5}
\plotone{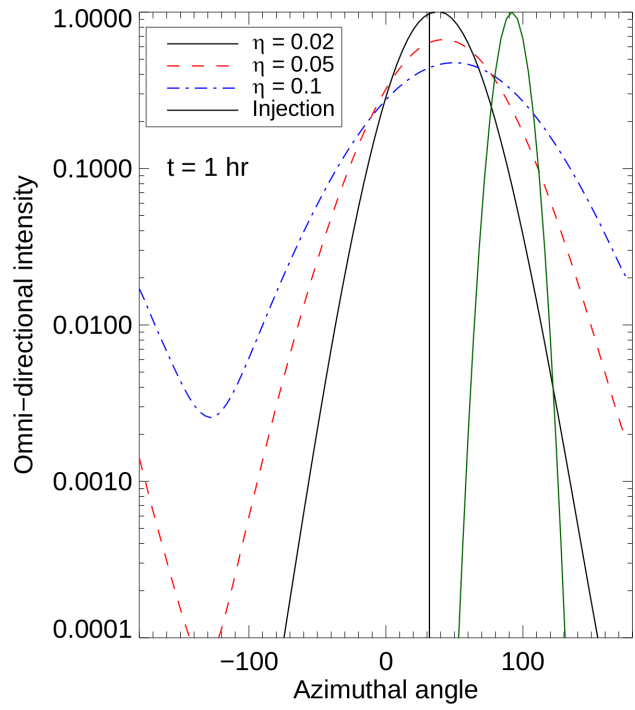}
\caption{{Similar to Fig. \ref{fig3}, but this time at $t=1$ hr, and for three different choices of $\eta$}.\label{fig7}}
\end{figure}

To more explicitly show the anti-symmetrical nature of the fluxes, Fig. \ref{fig7} gives the intensity as a function of $\phi$ at $t=1$ hr and $r=1$ AU. The figure is similar to Fig. \ref{fig3}. Three solutions are shown, corresponding to different assumptions of $\eta$, as indicated in the legend. It is clear that the distributions are neither symmetrical about their maxima (again, enhanced towards the west), nor does the azimuthal position of the maximum flux at 1 AU occur at the position of optimal magnetic connectivity. The latter quantity is also shifted towards the west, while this shift is larger for larger values of $\eta$.

\section{Towards Observables}

\begin{figure}
\epsscale{.99}
\plotone{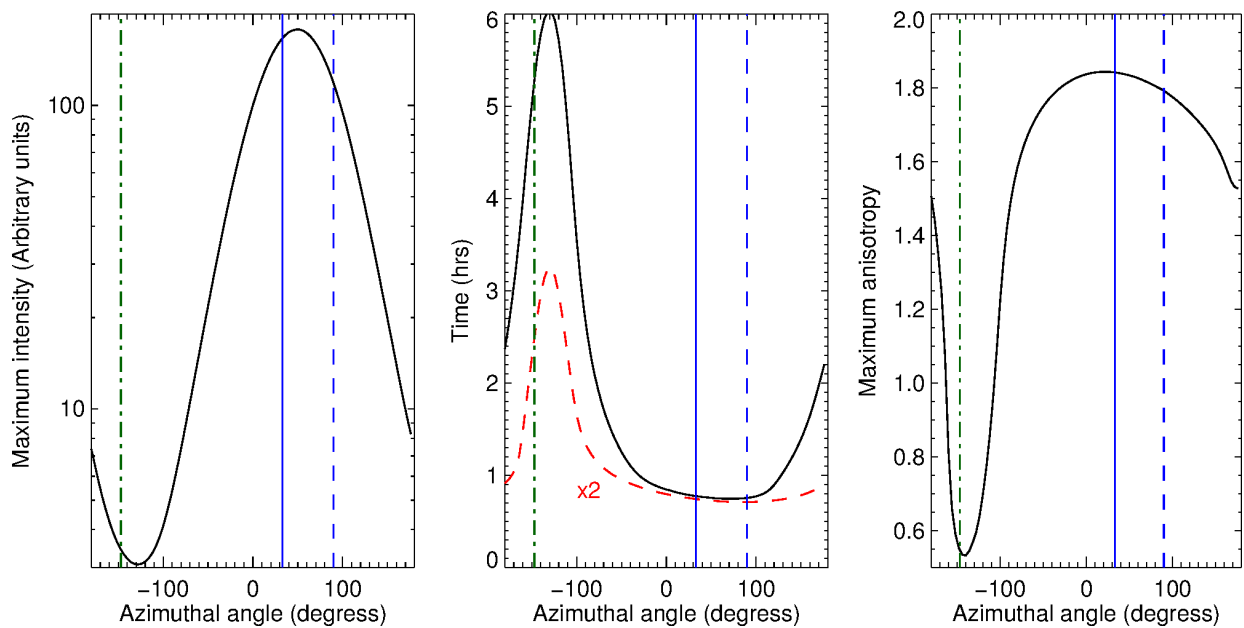}
\caption{{The figure shows, from left to right, the following quantities as a function of azimuthal angle at $r=1$ AU: The maximum intensity, both the time of maximum intensity and the onset time (note that the onset time is multiplied by a factor of 2) and the maximum anisotropy. The dashed blue line indicates where the injection function obtains its maximum value at the inner boundary, the solid blue line the position of optimal magnetic connectivity at 1 AU and the dash-dotted line the position of worst magnetic connectivity}.\label{fig8}}
\end{figure}

Here, the results shown in the previous section are presented in terms of observable quantities, i.e. in terms of observables used by the experimental community \citep[see e.g.][]{dressing2014}. These results, assuming $\eta=0.1$, are shown in Fig. \ref{fig8}, again as a function of azimuthal angle at $r=1$ AU. In this graph, the solid vertical line indicates the position of optimal magnetic connectivity at 1 AU to the source, the green dash-dotted line the angle of worst magnetic connection ($180^{\circ}$ away from the best magnetic connection point) and the dashed black line the position where the injection function reaches a maximum at the inner boundary. In the left panel, the maximum intensity is shown, that is, the maximum intensity for all times, recorded for each $\phi$. As noted previously, the maximum of this distribution is shifted towards the west of best magnetic connection. The middle panel shows the time of maximum (solid curve), which is defined as the time when the maximum intensity is reached at each $\phi$. Also shown in the middle panel is the onset time at that $\phi$ (multiplied by a factor of 2; dashed red line). This last quantity is difficult to define, and for the purposes of this study, it is defined as: That time, at a given $(r,\phi)$ when the SEP distribution reaches 1/100 of its global maximum value (i.e. the maximum of all $\phi$'s at all times). A similar approach was followed by \citet{wangqin2014}. In this way, we mimic in the model a certain background level as experienced in the experimental case. Both of these time scales are roughly anti-correlated with the maximum intensity; a higher maximum intensity usually corresponds to a shorter propagation time and hence, a shorter onset time and a shorter time needed to reach this maximum intensity value. A correlation between the time of maximum and the onset time is also evident although this relationship may be non-linear. The right panel shows the maximum anisotropy, generally occurring close to the onset time. The maximum anisotropy is again anti-correlated with the propagation time scales. It is believed that SEPs that take longer to reach e.g. 1 AU, must experience more (pitch-angle and perpendicular) diffusion, and hence, the distribution of these particles become increasingly isotropic.

\section{Discussion}
\label{Sec:discuss}

In this study, we have constructed a numerical SEP transport model and examined the effect of perpendicular diffusion on the resulting intensities. For illustrative purposes, simplified transport parameters were implemented, while, in future, more realistic coefficients will be used \citep[as in, e.g.,][]{hewang2012} and the results will be compared directly with observations. The qualitative conclusions presented in this study are, however, not expected to change.

We have shown that different functional forms (pitch-angle dependencies) of $D_{\perp}(\mu)$ can lead to rather different SEP intensities at Earth, even if the resulting perpendicular mean free path is the same. Generally, perpendicular diffusion coefficients which have a maximum near $\mu=1$, lead to the most effective perpendicular transport of SEPs because of the highly anisotropic SEP distribution near the Sun. The anisotropic nature of the distribution is caused by effective focussing between the Sun and Earth. The FLRW coefficient is found to be the most effective one and leads to the broadest longitudinal distribution of SEPs. It must however be noted that while the FLRW is known to be very effective (compared to the coefficients derived from other diffusion theories), it may overestimate the perpendicular diffusion process \citep[][]{kobus}. The pitch-angle dependence of $D_{\perp}$ has been neglected as a significant transport parameter in the past, but our results indicate that more care must be taken regarding its choice.

By calculating SEP intensities along the Earth's orbit (as a function of longitude at 1 AU), it was shown that the resulting distribution is asymmetrical in terms of longitude, with the intensities enhanced towards the west of optimal magnetic connectivity to the acceleration region (i.e. the source). This was demonstrated to be due to the geometry of the HMF, where perpendicular diffusion becomes increasingly directed in the radial direction at larger radial distances. Moreover, it was shown that, because of the non-constant transport parameters (i.e. their spatial derivatives are non-zero in the global coordinate frame), the maximum intensity of the SEP distribution may also be shifted towards the west of the line of best magnetic connection. A careful comparison to observations may, in future, quantify this effect in more detail, although our results are in qualitative agreement with the measurements discussed by, amongst others, \citet{richardsonetal2014}.

Observable quantities, including the maximum intensity, the time of maximum intensity, the onset time and the maximum anisotropy were calculated for different degrees of magnetic connectivity to the source (i.e. at different longitudes). The maximum intensity and maximum anisotropy seem to be correlated, while both are anti-correlated to the time of maximum and the onset time. Although \citet{wibberenzane2006} found that the maximum intensity is well correlated with the level of magnetic connectivity to the source region, as illustrated in this work, they found no clear azimuthal dependence for the time of maximum. A more detailed study by \citet{richardsonetal2014} did however find that both the time of maximum intensity and the onset time reach their minimum values near $\phi$-values of best magnetic connectivity. Moreover, \citet{richardsonetal2014} also found that both of these quantities (as well as the peak intensity) seems to be shifted towards the west of best connection; consistent with the modelled solutions presented here. The modelled azimuthal dependence of the maximum anisotropy seems, furthermore, to be consistent with the results of \citet{dressing2014}. 

Lastly, we discuss the effect that some choices of $D_{\perp}(\mu)$ do not seem to preserve causality.   Although it is well known that all diffusion equations exhibit this behaviour -- a delta function is, for example, instantaneously transformed into a Gaussian distribution \citep[e.g.][]{causaldiff}; refer also to the telegraph equation \citep[][]{telegraph} -- the effect discussed in this paper is due to more fundamental considerations as discussed below. Consider the TPE in the limiting case of $\mu=\pm1$

\begin{equation}
\mu=\pm1 : \frac{\partial f}{\partial t} = \mp v\frac{\partial f}{\partial z} + \frac{\partial}{\partial x} \left( D_{\perp} \frac{\partial f}{\partial x} \right)
\end{equation}

in a HMF aligned coordinate system with $\hat{z}\cdot\hat{b}=1$ and $\hat{x} \cdot \hat{b}=0$. This form of the TPE follows from the fact that both the pitch-angle diffusion coefficient and the focusing terms becomes zero at $\mu=\pm 1$. In a given time step $\Delta t$, a particle will move by $\Delta z = \pm v\Delta t$ in the $\hat{z}$-direction, while, simultaneously diffusing by $\Delta x$ along $\hat{x}$. Note that for choices of $D_{\perp}(\mu=\pm 1) \neq 0$ (for example the FLRW coefficient), $\Delta x \neq 0$. The total displacement of such a particle is then

\begin{equation}
\Delta s = \sqrt{(\Delta z)^2 + (\Delta x)^2} > v \Delta t,
\end{equation}

so that $\Delta s/\Delta t > v$. This means that when $D_{\perp} (\mu=\pm 1) \neq 0$, SEPs propagate faster than their actual speed allows; the addition of perpendicular diffusion causes an artificial acceleration of the particles. Although this is most evident at $\mu=\pm1$, this may also occur at other values of $\mu$. Although we have illustrated this possible inconsistency here, we are neither sure why it exists, nor do we know how to overcome this difficulty (if the latter is actually needed). It is interesting to note that the FLRW diffusion process, as implemented by \citet{Laitinen}, where a stochastically varying $\hat{b}$ is specified, does not violate causality. The inconsistency between these two approaches is indeed worrying and needs future investigation. 

\acknowledgments

RDS acknowledges the partial financial support of the South African National Research Foundation (NRF). This research was partially funded by the Alexander von Humboldt Foundation. The authors acknowledge informative discussions with Drs. H.-Q. He and N. Dresing regarding the manuscript. The work also benefited from discussions at the team meeting ``Superdiffusive Transport in Space Plasmas and its Influence on Energetic Particle Acceleration and Propagation'', supported by the International Space Science Institute (ISSI) in Bern, Switzerland.

\appendix

\section{Numerical Aspects of the Transport Model}
\label{Sec:appen1}

As we are solving Eq. \ref{Eq:TPE} in spherical spatial coordinates, all transport quantities, as per usual specified in a local HMF aligned coordinates system, must be transformed to global spherical coordinates. These transformations are briefly illustrated below, whereafter some important aspects of the numerical scheme are discussed.

The standard Parker HMF is given by

\begin{equation}
\vec{B}(r,\theta) = \frac{B_0r_0^2}{r^2} \left( \hat{r} - \tan \Psi \hat{\phi} \right)
\end{equation}

where $B_0$ is some reference value at $r_0$. The magnitude of the HMF however never enter any calculations here, and only the geometry is of importance. The HMF spiral angle ($\Psi$, the angle between the HMF and the radial direction) is defined by

\begin{equation}
\tan \Psi = \frac{\Omega r \sin \theta}{V_{sw}}
\end{equation}

with $\Omega$ the angular rotation speed of the Sun and $V_{sw}=400$ km.s$^{-1}$ the solar wind speed. As the model is limited to the ecliptic regions of the heliosphere, $\sin \theta = 1$ is assumed throughout. Moreover, since the Parker HMF is independent of $\phi$, all transport quantities are also assumed to be so. 

\begin{figure}
\epsscale{.50}
\plotone{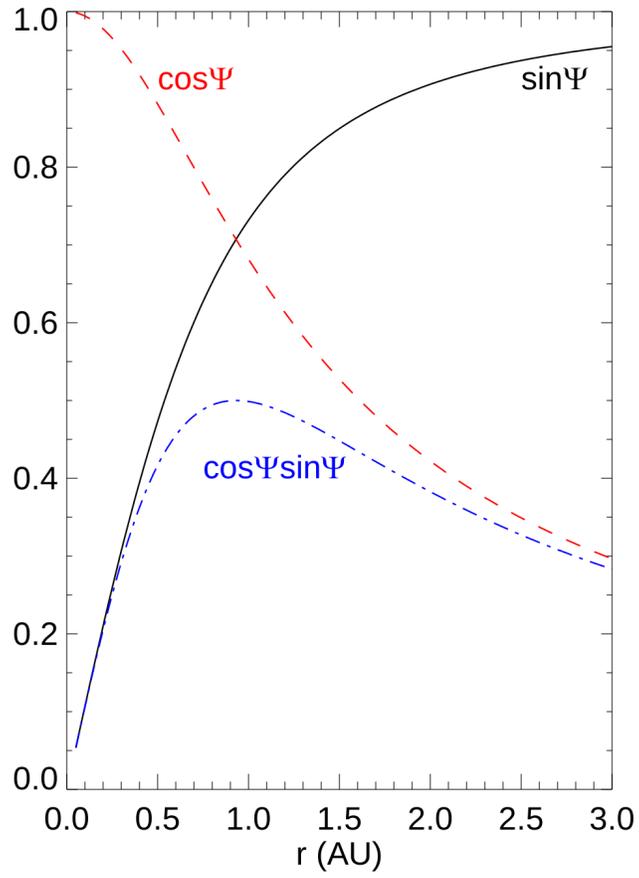}
\caption{{Showing the radial dependence of $\cos \Psi$, $\sin \Psi$ and $\cos\Psi \sin \Psi$, calculated for a Parker HMF geometry}.\label{figA1}}
\end{figure}

With this definition,

\begin{equation}
\hat{b} = \cos \Psi \hat{r} - \sin \Psi \hat{\phi}
\end{equation}

which determines the streaming direction in Eq. \ref{Eq:TPE}, while the diffusion tensor takes the form (see also the discussion by \citet{fred2012})

\begin{equation}
\mathbf{D}_{\perp} =
\left( \begin{array}{cc}
D_{\perp}^{(rr)}  & D_{\perp}^{(r\phi)}\\
D_{\perp}^{(\phi r)}  & D_{\perp}^{(\phi\phi)} \end{array} \right) =
 \left( \begin{array}{cc}
D_{\perp} \sin^2 \Psi  & D_{\perp} \sin \Psi \cos \Psi\\
D_{\perp} \sin \Psi \cos \Psi  & D_{\perp}  \cos^2 \Psi \end{array} \right) 
\end{equation}

where $D_{\perp}$ is the perpendicular diffusion coefficient specified in the local HMF aligned coordinate system. Also note that $D_{\perp}^{(\phi r)} =D_{\perp}^{( r \phi )} $. As an illustration, the values of $\cos \Psi$, $\sin \Psi$ and $\cos \Psi \sin \Psi$ are shown in Fig. \ref{figA1} as a function of radial distance.

The TPE in spherical coordinates then becomes

\begin{eqnarray}
\frac{\partial f}{\partial t} &+& \overbrace{\frac{1}{r^2} \frac{\partial}{\partial r} \left( \mu v \cos \Psi \left(r^2f \right) \right)}^{\mathrm{streaming \ in \ }r} 
+ \overbrace{\frac{\partial}{\partial \phi} \left(- \frac{\mu v \sin \Psi}{r} f \right)}^{\mathrm{streaming \ in  \ }\phi}
+ \overbrace{\frac{\partial}{\partial \mu} \left( \frac{1-\mu^2}{2L} vf \right)}^\mathrm{focusing}\\
&=& \overbrace{\frac{1}{r^2} \left( r^2 D_{\perp}^{(rr)} \right)\frac{\partial f}{\partial r} + \frac{D_{\perp}^{(r\phi)}}{r} \frac{\partial ^2 f}{\partial r \partial \phi}   + D_{\perp}^{(rr)} \frac{\partial^2 f}{\partial r^2}}^{\mathrm{diffusion \ in \ } r} + \overbrace{\frac{\partial}{\partial \mu} \left(D_{\mu\mu} \frac{\partial f}{\partial \mu} \right)}^{\mathrm{diffusion \ in \ }\mu} \\
&+& \overbrace{\frac{1}{r^2} \frac{\partial}{\partial r} \left( r D_{\perp}^{(\phi r)} \right) \frac{\partial f}{\partial \phi} + \frac{D_{\perp}^{(\phi r)}}{r} \frac{\partial ^2f}{\partial r \partial \phi} + \frac{D_{\perp}^{(\phi \phi)}}{r^2} \frac{\partial^2 f}{\partial \phi^2}}^{\mathrm{diffusion \ in \ } \phi}
\end{eqnarray}

which is the equation to be solved by applying a suitable numerical scheme.

A numerical solution of the equation above requires some careful consideration: If the advection terms (streaming and focusing) would dominate, the equation can become hyperbolic in nature, while, when diffusion dominates, it may become increasingly parabolic. It is, therefore, unlikely (if not impossible) that a single numerical scheme could handle this equation. We have opted to solve the TPE by applying the operator splitting technique \citep[see amongst others][]{marchuk}. A similar treatment was also considered by \citet{hatzky} and \citet{lampas}. Here, the TPE is split along both spatial and pitch-angle coordinates and along first and second order terms, i.e. the differential operator becomes

\begin{equation}
\mathcal{L} = \sum_{i = 1}^{3} \mathcal{L}^{\mathrm{advection}}_{i} + \sum_{i = 1}^{3} \mathcal{L}^{\mathrm{diffusion}}_{i}
\end{equation}

with $i \in \left\{ r,\phi,\mu \right\}$, to give six one dimensional differential equations (three of which are of first order and three of second order). As an example, the equations for the $\mu$ dimension becomes

\begin{eqnarray}
\frac{1}{6}\frac{\partial f}{\partial t'} &=&  \frac{\partial f}{\partial t} = - \frac{\partial}{\partial \mu} \left( \frac{1-\mu^2}{2L} vf \right)\\
\frac{1}{6}\frac{\partial f}{\partial t'} &=& \frac{\partial f}{\partial t} = \frac{\partial D_{\mu\mu}}{\partial \mu}\frac{\partial f}{\partial \mu} + D_{\mu\mu} \frac{\partial f^2}{\partial \mu^2}
\end{eqnarray}

where $dt' = dt/6$. It can be shown that for this operator splitting algorithm, the accuracy is only to first order, $\Delta f \sim \mathcal{O} (\Delta t)$. The upside is, however, that different numerical schemes (and even different boundary conditions) can be applied to each resulting equation. The diffusion equations are solved by a simple explicit time-forward central difference scheme with accuracy $\Delta f \sim \mathcal{O}(\Delta t) + \mathcal{O}(\Delta x)^2$. A more accurate (in time) method is superfluous, as the temporal accuracy is already limited by the splitting of the differential operators. For the advection equations, an upwind scheme is employed \citep[see e.g.][]{tracpen}, together with the \citet{vanleer1974} flux limiter, to give $\Delta f \sim \mathcal{O}(\Delta t^2) + \mathcal{O}(\Delta x^2)$. As a whole, the numerical scheme has a numerical accuracy of $\Delta f \sim \mathcal{O}(\Delta t) + \mathcal{O}(\Delta x^2)$.

\begin{figure}
\epsscale{.50}
\plotone{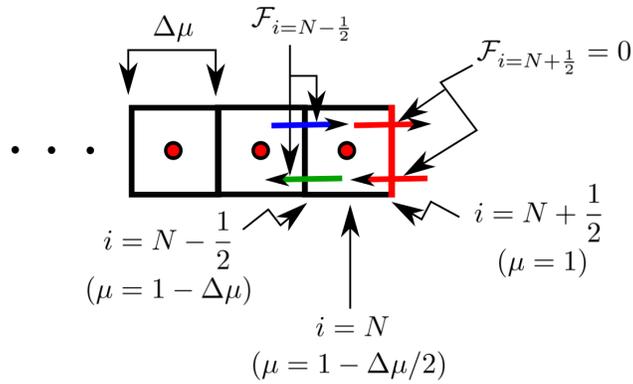}
\caption{{The numerical set-up used to solve the $\mu$ advection and diffusion equations. The problem is to find the appropriate boundary conditions for $f^t_{i=N}$. Here, this limitation is overcome by examining the fluxes into (blue arrow) and out of (green arrow) the last grid cell. Note that the flux through the cell face $i=N+1/2 (\Rightarrow \mu=1) $ (red arrows) is zero due to the adopted choices of the transport parameters}.\label{figA2}}
\end{figure}

The boundary conditions for $r$ and $\phi$ are straight forward: An injection function is specified at $r_0=0.05$ AU and an absorbing condition at $r_b=3$ AU, while periodic boundary conditions are used for $\phi$. The boundary conditions for $\mu$, however, require careful consideration because an incorrect choice for these can easily lead to violation of particle conservation. Fig. \ref{figA2} shows a portion of the $\mu$ grid near $\mu=1$ illustrating the approach followed here: $f_i$ is specified at the cell centres, e.g. $i=N$ (which is located at $\mu=1-\Delta \mu/2$), while the cell faces are located at $i=N\pm 1/2$. To find suitable boundary conditions for $f_i$, we can examine the fluxes entering (blue arrow in the figure) and exiting (green arrow in the figure) this computational cell. Note that, because the pitch-angle diffusion and focusing terms are both zero at $\mu=\pm1$, the flux through cell face $i=N+1/2$ (located at $\mu = 1$) is always zero (the red arrows in the figure), so that we may compute

\begin{equation}
f_{i=N}^{t+\Delta t} = f_{i=N}^{t} + \frac{\Delta t}{\Delta \mu} \mathcal{F}^t_{i=N-1/2}
\end{equation}

where $\mathcal{F}^t_{i=N-1/2}$ is the (advective or diffusive) flux entering or leaving the last cell. For the $\mu$-advection equation, this is

\begin{equation}
\mathcal{F}^{\mathrm{advective}}_{i=N-1/2} = \left. \frac{v(1-\mu_i)^2)}{2L}  f_{i}^t\right|_{i=N-1},
\end{equation}

while, for the pitch-angle diffusion term, it becomes

\begin{equation}
\mathcal{F}^{\mathrm{diffusive}}_{i=N-1/2} = -\tilde{D}_{\mu\mu} \left. \frac{\partial f}{\partial \mu} \right|_{i=N-1/2},
\end{equation}

where 

\begin{equation}
\tilde{D}_{\mu\mu} \approx \frac{1}{2} \left\{ D_{\mu\mu,i=N} + D_{\mu\mu,i=N-1} \right\}
\end{equation}

and

\begin{equation}
\left. \frac{\partial f}{\partial \mu} \right|_{i=N-1/2} \approx \frac{1}{\Delta \mu} \left\{ f_{i=N}^t - f^t_{i=N-1} \right\}.
\end{equation}

Similar treatments of the fluxes have been implemented in the past by e.g. \citet{ngwong1979} and \citet{koteatel1982}.

\begin{figure}
\epsscale{.99}
\plotone{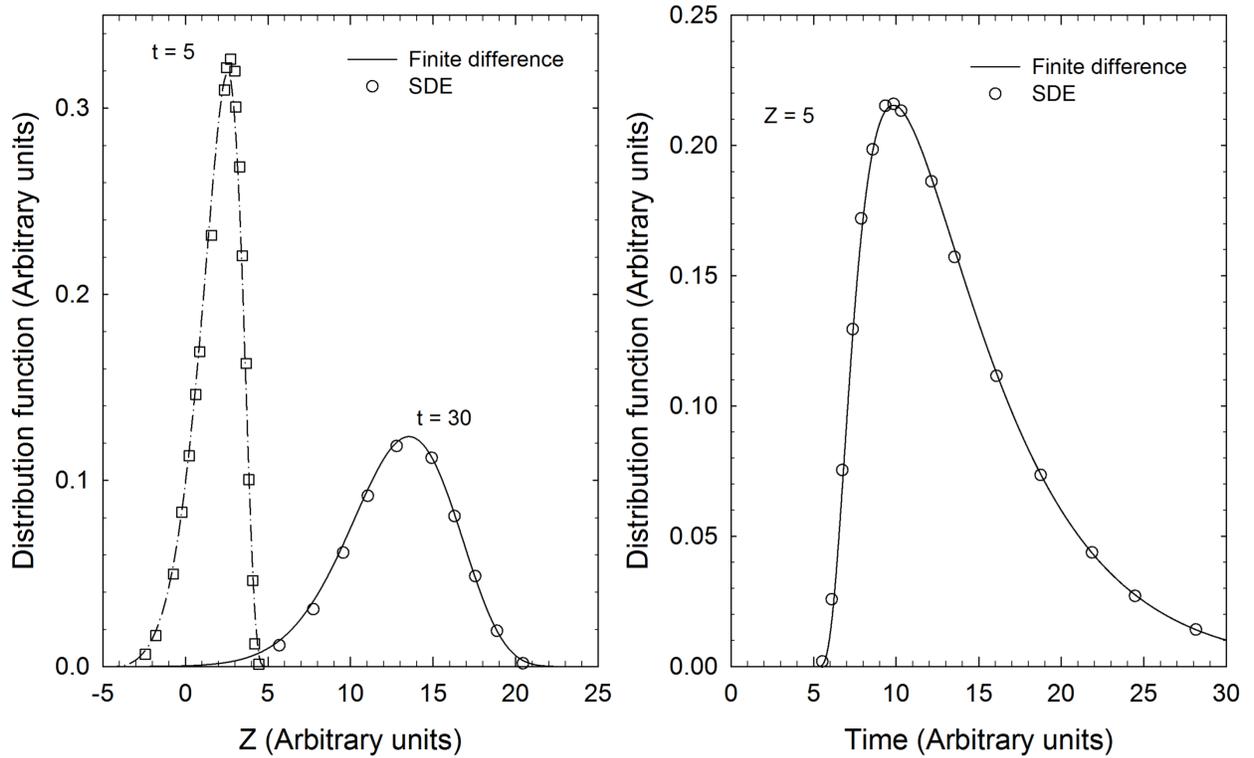}
\caption{{Comparing the model discussed in this work (solid lines), with the SDE based model of \citet{fred2014} (symbols). We are able to reproduce their results (see their Figs. 3 and 5) very accurately}.\label{figA3}}
\end{figure}

Fig. \ref{figA3} shows an example of the benchmarking studies performed on the present model. Here, we solve the \citet{roelof1969} equation with this numerical scheme (solid lines), and compare the results to the stochastic differential equation (SDE) based model of \citet{fred2014} using the same transport parameters (see their Figs. 3 and 5). The modelled results vindicate the modelling approach outlined in this paper, and more importantly, because the SDE model conserves particles by construction, we are confident the same can be said of the present model.


\clearpage


\clearpage

\clearpage

\clearpage

\clearpage

\clearpage

\clearpage

\clearpage

\clearpage

\clearpage

\clearpage

\clearpage

\clearpage

\end{document}